\documentclass[sigconf,screen]{acmart}

\AtBeginDocument{%
  }

\usepackage{todonotes}
\usepackage{paralist}
\usepackage{balance}
\usepackage{cleveref}
\usepackage[utf8]{inputenc}
\usepackage{pgfplots}\pgfplotsset{compat=1.17}
\usepackage{xspace}
\newcommand{\ie}{\textit{i.e.,}\xspace}
\newcommand{\eg}{\textit{e.g.,}\xspace}

\newcommand{\etal}{\textit{et al.}\xspace}

\begin{document}


\title{Augmenting Software Bills of Materials with Software Vulnerability Description: A Preliminary Study on GitHub}

\author{Davide Fucci}
\email{davide.fucci@bth.se}
\orcid{0000-0002-0679-4361}
\affiliation{%
  \institution{Blekinge Institute of Technology}
  \city{Karlskrona}
  \country{Sweden}
}

\author{Massimiliano Di Penta}
\orcid{0000-0002-0340-9747}
\affiliation{%
  \institution{University of Sannio}
  \city{Benevento}
  \country{Italy}}
\email{dipenta@unisannio.it}

\author{Simone Romano}
\email{siromano@unisa.it}
\orcid{0000-0003-4880-3622}
\affiliation{%
  \institution{University of Salerno}
  \city{Fisciano}
  \country{Italy}
}

\author{Giuseppe Scanniello}
\email{gscanniello@unisa.it}
\orcid{0000-0003-0024-7508}
\affiliation{%
 \institution{University of Salerno}
 \city{Fisciano}
 \country{Italy}}


\begin{abstract}
\textit{Software Bills of Material} (\textit{SBOMs}) are becoming a consolidated---and often enforced by governmental regulations---way to describe software composition. However, based on recent studies, SBOMs suffer from limited support for their consumption, and lack information beyond simple dependencies, especially regarding software vulnerabilities. 
This paper reports the results of a preliminary study in which we augmented SBOMs of 40 open-source projects with information about \textit{Common Vulnerabilities and Exposures} (\textit{CVE}) exposed by project dependencies. Our augmented SBOMs have been evaluated by submitting pull requests and by asking project owners to answer a survey. Although in most cases, augmented SBOMs were not directly accepted because owners required a continuous SBOM update, the received feedback shows the usefulness of the suggested SBOM augmentation.
\end{abstract}


\maketitle

\section{Introduction}
The transparency of software products in terms of their components, known as \textit{Software Composition Analysis} (\textit{SCA}) \cite{sca1,ChenSS020}, is desirable for multiple reasons. On the one side, this allows software consumers to know whether the software contains components originating from undesired providers. For example, a government might avoid using, for their applications, software originating from hostile countries. Second, the analysis of software components helps ensure the software is not violating licensing---\eg by containing components with an incompatible license, or copyrights. Last, and often more importantly, software composition allows knowing whether a piece of software depends on a vulnerable component.

A pretty-consolidated, and machine-readable way to describe software composition and, in general, detailing different pieces of information concerning supply-chain relationships, is a \textit{Software Bills of Material} (\textit{SBOM}). SBOM adoption is, in recent years, not only motivated by the aforementioned needs but also by governmental regulations, such as the 2021 United States Executive Order 14028 \cite{USA:ExecutiveOrder} or the more recent European Union \textit{Cyber Resilience~Act} (\textit{CRA})~\cite{EU:CRA}.

Several researchers have empirically investigated the needs, adoption, and challenges of utilizing SBOMs in software development and distribution \cite{Linux:2022,Xia:2023,Stalnaker:2023,Kloeg:2024,Chaora:2023,Nocera:2024}. 
On the one hand, there is a general agreement on the usefulness of SBOMs, and there is an increasing availability of tool support for SBOM generation, although the latter is generally limited to analyzing and documenting package-level dependencies from manifest/lock files. On the other hand, the conducted studies pointed out insufficient support in terms of tools for SBOM  consumption, and a limited presence of information necessary for software vulnerability assessment. 
More specifically, while in principle existing SBOM formats allow explicitly documenting software vulnerabilities, SBOM files (or simply SBOMs, from here onwards) generated by existing tools (\eg CycloneDX tools or the GitHub SBOM generator) merely document the (directly or indirectly) used components and their versions, and a vulnerability assessment remains a task to be conducted by the SBOM consumer using other software tools. It appears to be of significant importance to ask the following question:

\begin{quote}
\emph{To what extent would it be useful to augment SBOMs with information about components' vulnerabilities?}
\end{quote}

Certainly, one may wonder whether explicitly documenting the presence of vulnerable components might end up sabotaging the diffusion and success of a software product, and, therefore, something software manufacturers might avoid. At the same time, documenting potential vulnerabilities may represent a desirable property from the consumer perspective---considering also that a dependency vulnerability can be protected on the client's side---and therefore software manufacturers might be willing to include it, having suitable tools and guidelines for that purpose. To summarize, answering the research question above could be pivotal for bridging the gap between static software transparency tools (\eg SBOMs) and dynamic vulnerability-related information, ultimately contributing to safer and more resilient software.

To provide initial evidence and clarify this doubt, in this paper, we report the result of a preliminary empirical study in which we augmented the SBOMs of 40 existing open-source projects hosted on GitHub, and investigated whether project owners were willing to leverage the augmented SBOMs.  
In this regard, we first searched for open-source projects containing SBOM files in their GitHub repository, ranked, and filtered them based on stars and development activity (number of contributors and commits).
Then, we analyzed their dependencies using the \texttt{osv-scanner}~\cite{osvscanner} tool to identify those depending on a component exposing at least one cybersecurity vulnerability in \textit{Common Vulnerabilities and Exposures} (\textit{CVE}) databases. Then, we generated files in the \textit{VEX} (\textit{Vulnerability Exploitability eXchange}) format---an industry standard for describing software vulnerabilities and their exploitability \cite{vex}---to complement the existing SBOM with vulnerability-related information (\ie a software vulnerability description). Last but not least, we opened a Pull Request (PR) suggesting augmenting the existing SBOM with the generated VEX document. Within the PR, we also asked the project developers/owners to answer a (short) survey related to the usefulness of vulnerability-augmented SBOMs.

The feedback provided by project contributors who participated in the study, by either engaging in PR discussions or by answering a survey we attached to the PRs, confirmed the value of augmenting SBOMs with vulnerability-related information. At the same time, they expressed the need for a continuous generation of such pieces of information, possibly through tools integrated into the project's Continuous Integration (CI) pipeline.

All data, code, and materials necessary to replicate the findings of this study are available.\footnote{\url{https://github.com/SESAM-project/SBOM-VEX-acceptance}.} 

\section{Background and Related Work}
This section provides background notions and literature analysis on SBOMs and their adoptions, needs, and challenges. Instead, we do not focus our related work analysis on software security assessment, because we are not interested in understanding how vulnerability analysis is performed, but, rather, the extent to which vulnerability-related information can be used to augment SBOMs.

An SBOM is a formal machine-readable document that contains the details and supply-chain relationships of open-source and proprietary components used in building a software product~\cite{CISA:framing}. Public administrations have been pushing software manufacturers to use SBOMs in both the United States and the European Union.
Indeed, in 2021, the United States Federal Government, per President Biden's Executive Order 14028, laid down that any manufacturer releasing a software product to a federal agency must provide an SBOM of the released product~\cite{USA:ExecutiveOrder}. In 2022, the European Commission proposed the CRA~\cite{EU:CRA}, which entered into force in 2024. This regulation forces software manufacturers operating in the European Union to document vulnerabilities and components of their software products with the support of SBOMs. 

There are currently two main SBOM standards, namely \textit{SPDX} (\textit{Software Product Data Exchange}) and \textit{CycloneDX}~\cite{Xia:2023}, which support the baseline component information for SBOMs~\cite{CISA:framing}. Although SBOM standards allow adding vulnerability information within an SBOM, it is recommended to use a separate document linked to the SBOM, called VEX, to store vulnerability information~\cite{CISA:framing}. In particular, VEX documents are used to convey the status (\ie not affected, affected, fixed, or under investigation) of vulnerabilities within the components of a software product. SBOMs augmented with VEX documents should help software manufacturers, users, and other defenders more quickly and accurately assess the risks due to vulnerable components, which are often hidden behind opaque supply chain relationships~\cite{CISA:framing}. 

After President Biden's Executive Order 14028~\cite{USA:ExecutiveOrder}, there has been an increasing interest in SBOMs. In 2022, the Linux Foundation~\cite{Linux:2022} surveyed 412 worldwide organizations to understand SBOM's readiness. The results of this survey show gaps in familiarity with, production planning for, and consumption of SBOMs. Xia \etal~\cite{Xia:2023} conducted 17 interviews followed by a survey with 65 respondents to understand how practitioners perceive SBOMs. One of the main findings is that improving transparency and visibility into software products is deemed the main benefit of SBOMs. Stalknaker \etal~\cite{Stalnaker:2023} conducted a study combining surveys and interviews with practitioners to investigate how SBOMs are adopted by five groups of stakeholders. Stalknaker \etal~observed that the adoption of SBOMs
is still low, although practitioners have been using them in a variety of use cases---\eg 
software licensing or security assessment. Moreover, the study pointed out some limitations of the current technology and practice for what concerns SBOM consumption, in particular, related to security assessment.
Nocera \etal~\cite{Nocera:2023} analyzed the extent to which open-source projects hosted on GitHub use SBOM generation tools and publish SBOMs. The study results indicate that, while SBOM adoption is still low, it is trending upward, and SBOMs are published in only 46\% of the projects using SBOM generation tools. Kloeg \etal~\cite{Kloeg:2024} interviewed groups of stakeholders involved in SBOM production and consumption to delve into the reasons for low SBOM adoption. One of the main reasons is the lack of knowledge or expertise about SBOMs. Chaora \etal~\cite{Chaora:2023} analyzed the talks on SBOMs that took place in public SBOM community engagement meetings. To promote SBOM adoption, the authors foresaw regulations and a public-recognized agency that audits and validates SBOMs submitted to a common infrastructure. Finally, Nocera \etal~\cite{Nocera:2024} interviewed 10 practitioners working in the Italian software industry about SBOMs. The study results indicate that SBOM adoption is low, yet the attention of the Italian software industry to software supply chain-related challenges is high. 

To summarize, previous research has pointed out the importance of SBOM usage but also pointed out current limitations related to the technology and SBOM content. Related to the goal of our work, the latter lacks vulnerability-related information.

\begin{figure*}[t]
  \centering
  \includegraphics[width=\linewidth]{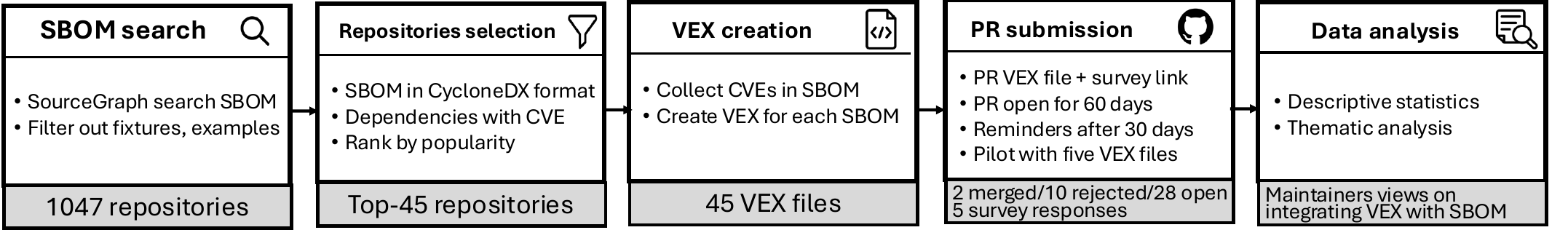}
  \caption{Methodology overview.}
  \label{fig:process}
  \vspace{-2mm}
\end{figure*}

\section{Study Methodology}
As stated in the introduction, the \emph{goal} of this study is to provide initial evidence on the perceived usefulness of augmenting SBOMs with vulnerability-related information. The \emph{context} consists of 40 pull requests from active open-source Java projects hosted on GitHub.
To address this goal, we focus on a more specific Research Question (RQ):
\begin{quote}
{\itshape What do open-source maintainers think about integrating VEX into their existing SBOMs?}
\end{quote}
\noindent The study of this RQ instantiates our general goal into a process aimed at collecting feedback from maintainers/developers after showing them SBOMs augmented with vulnerability information in the VEX format.

Our methodology consists of the steps presented in~\Cref{fig:process}.
We start by searching and filtering SBOMs in GitHub leveraging the SourceGraph \cite{sourcegraph} code search engine.
To that end, we used regular expressions tailored to the two most popular SBOM formats: SPDX~\cite{SpdxDraft:2010} and CycloneDX \cite{cyclonedx}. 
These regular expressions targeted mandatory attributes (\eg the string \texttt{SPDXID} for SPDX and \texttt{bomFormat} for CycloneDX) and specific file extensions (\eg \texttt{.yaml}, \texttt{.spdx}, \texttt{.xml}, and \texttt{.json}).
Furthermore, we excluded SBOM files intended as fixtures or used for testing purposes by employing regex expressions such as \texttt{*fixture*}, \texttt{example}, and \texttt{test}.
Additionally, SBOM files that did not declare any dependencies were filtered out, yielding 1,047 repositories containing relevant SBOM data.
It is important to mention that the goal of this preliminary screening was to create a relatively large dataset of open-source projects hosted on GitHub and have SBOMs available in their repositories. However, in the study presented in this paper, we restricted the analysis to projects adopting the CycloneDX format only. The reason for this choice was practical and due to the adopted tool (\texttt{vexctl}) to generate VEX files, which did not support the embedded format to describe vulnerabilities required by the SPDX format. This choice reduced the number of studied repositories to 270.

Subsequently, we enriched repositories whose dependencies in the SBOM had at least one CVE according to the \texttt{osv-scanner}~\cite{osvscanner}.
We selected this tool as it aggregates vulnerabilities from 25 different ecosystems (\eg \texttt{npm}, \texttt{crates.io}) and databases (\eg GitHub Security Advisory, RustSect) and provides a human and machine-readable data format to describe vulnerabilities.\footnote{The vulnerability scan was run on 2024/10/03.}
Finally, using the GitHub API, we obtained, for each of the remaining 117 repositories, the number of stars, contributors, and total commits.
The latter were used to rank repositories by popularity, from which we selected the top 45 for further analysis. 

We downloaded all SBOM files present in these repositories and extracted information about vulnerabilities associated with the dependencies declared in each SBOM using \texttt{osv-scanner}.
Based on the identified vulnerabilities, we generated a VEX file for each SBOM using the \texttt{vexctl}~\cite{vexctl} tool.
We selected this tool as it is one of the few existing solutions to support the generation of VEX starting from SBOM and CVE information.

Next, we opened a PR in each selected repository, committing the VEX file.
The PR message included a link to an online survey, inviting respondents to rate their agreement on a 5-point Likert scale~\cite{Oppenheim:1992} about the perceived usefulness of using VEX to augment SBOMs with vulnerability information.
Moreover, the respondents could provide qualitative feedback explaining their responses.

We left the PRs open for 60 days, during which we monitored and collected survey responses as well as comments left directly on the PRs. After 30 days of inactivity (since the PR was opened), we sent reminders to the repository maintainers. The collected data were subsequently analyzed to assess the reception and perceived usefulness of integrating VEX into SBOM workflows.

Before the actual study, we performed a pilot of two weeks targeting five randomly selected repositories. Two PRs were accepted during the pilot, while the remaining three were not reviewed by the maintainers. One maintainer filled in the survey. Since no other feedback was provided during the pilot, we decided to carry on with the study. The pilot projects were excluded from the final analysis, resulting in a total of 40 projects being considered.

\section{Study Results}
We open a PR including the VEX file for an SBOM tracked in 40 GitHub repositories.
Out of these, two were merged, 10 were rejected and 28 did not receive a review by the repository maintainers (\ie open). 
The survey collected five complete responses from maintainers who confirmed that they had reviewed the PR.

\subsection{Analysis of Pull Request Comments}
The number of comments (excluding bot-generated comments and the reminder we sent to the maintainers) for PRs that were not closed during the time of the study ranged between zero and six. 
For rejected PRs, the number of comments in our discussion with the maintainers ranged between one and five.
Accepted PR discussion length ranged between zero and two comments.

The thematic analysis of comments regarding unmerged PRs revealed uncertainty surrounding the information contained in VEX files. Maintainers expressed willingness to adopt VEX only if provided with mechanisms for continuous (\eg triggered by SBOM generation) and automated updates (\eg integrated in CI pipelines).

Maintainers highlighted a critical limitation of the information reported in a VEX---\ie its tendency to become outdated, potentially misrepresenting the security status of project dependencies. Despite this, they recognized that the dynamic integration of VEX with SBOM could enhance overall security.

Another reason for rejecting PRs was the reliance on alternative tools, such as Dependabot~\cite{dependabot} and Grype~\cite{grype}, for scanning SBOMs for vulnerabilities. However, these tools do not effectively manage the lifecycle of identified vulnerabilities by tracking their state (\eg patched or unpatched) and rationale, nor do they facilitate communication of this information to users.

An additional challenge stems from the lack of standardization across SBOM formats and the varying approaches to integrating them with VEX. This ambiguity has deterred maintainers from adopting VEX or led them to delay its integration until more stable standards emerge. One discussion cited the forthcoming European Union CRA~\cite{EU:CRA} as a potential catalyst for advancing the combined use of SBOM and VEX.

Finally, while one maintainer acknowledged the importance of VEX, they emphasized its relevance primarily for internal use rather than for public projects.

\subsection{Analysis of Survey Results}
Among the five responses, one was from a maintainer who accepted the PR, two from maintainers who rejected it, and two from maintainers who had not yet decided after reviewing it.
One respondent strongly agreed that adding vulnerability information to dependencies listed in an SBOM file is beneficial, while two partially agreed. 
These respondents emphasized the importance of transparency in open-source project security, noting that VEX can contribute to achieving this goal. 
However, they also highlighted that the effectiveness of VEX depends on the quality of the CVE databases it relies on, with its utility potentially limited by the signal-to-noise ratio in publicly available vulnerability data.
The remaining two respondents disagreed, justifying their position by pointing to the availability of other tools and services capable of continuously scanning SBOM files. 

\subsection{Discussion}
The study participants expressed support for adopting VEX, provided it is automated and seamlessly integrated into project pipelines (\eg alongside SBOM creation and updates). However, some maintainers noted that VEX might be redundant since existing tools (\eg Grype) already report vulnerabilities. Nonetheless, such reports lack a standardized structure, making them non-interchangeable and unsuitable for tracking vulnerability lifecycles. More importantly, such reports cannot be easily integrated into SBOMs, the latter being considered particularly important for the studied projects (all the studied projects were adopting SBOMs already).

As a result, a key recommendation for tool providers is to support VEX format output as a \emph{lingua franca} for sharing information about vulnerabilities in project dependencies and to integrate the VEX document generation with the SBOM generation, for example within the project's CI pipeline.

Another significant challenge to VEX adoption is the lack of standardization. While SBOM has established two de facto standards—CycloneDX and SPDX—the former, supported by the OWASP Foundation, sponsors VEX format development, and the latter (SPDX v3.0) enables embedding vulnerability information. Such results are in line with concerns expressed in previous studies about the lack of a common standardization for SBOMs~\cite{Xia:2023,Stalnaker:2023}.

Meanwhile, legislators are increasingly requiring SBOMs and vulnerability management. However, the EU CRA, which became applicable in December 2027, does not mandate a specific SBOM format or the use of VEX for vulnerability management. To align with legislative demands, further standardization of VEX is essential.

\section{Threats to Validity}

In the following, we discuss the threats that might affect our results based on the validity schema by Wohlin \etal~\cite{Wohlin:2012}.

\noindent
\textit{Construct Validity.} The approach we used to detect the adoption of SBOMs might affect
the results. That is, the heuristic (based on the SourceGraph code search engine) to detect repositories may have missed some of them that used SBOMs. This threat is well recognized in studies similar to ours~\cite{Nocera:2023}. 

\noindent
\textit{Internal validity.} A possible threat to internal validity is self-selection bias because those who engaged in the PR discussion/survey could be developers more interested in adopting the proposed SBOM augmentation. This threat is partially mitigated by, at least, already focusing on projects adopting SBOMs, and, therefore, considering supply chain documentation important.

\noindent
\textit{Conclusion Validity.} Our study is exploratory and did not use inferential statistics to answer our RQ. Therefore, this kind of validity, which concerns the ability to draw the correct conclusions~\cite{Wohlin:2012}, is limited in our study. There could still be risks that the results are affected by: (i)~limited time (60 days) for contributors to review our PRs, (ii)~limited activity of contributors on the repositories suggesting scant interest of developers on the hosted projects, and (iii)~only maintainers interested in the topic of our study answered the survey on vulnerability-augmented SBOMs. Given the preliminary nature of our study, these threats to validity are not of significant concern.

\noindent
\textit{External Validity.} The most important threat to this kind of validity is that our study is limited to: {(i)}~open-source projects hosted on GitHub and {(ii)}~SBOM generation tools owned by CycloneDX. The second point does not represent an actual threat because we opened a PR suggesting augmenting the existing SBOM with the generated VEX and asked owners to answer a survey on vulnerability-augmented SBOMs. Given our objective, the consideration of CycloneDX (or any other SBOM standard) is not a primary concern.

\section{Conclusion and Future work}

In this paper, we presented the initial results of a preliminary empirical study where we enhanced the Software Bill of Materials (SBOMs) of 40 open-source projects by incorporating information about Common Vulnerabilities and Exposures (CVEs) associated with project dependencies. The augmented SBOMs have been evaluated by submitting PRs and by asking project owners to answer a survey. In most cases, our augmented SBOMs were not directly accepted because developers required a continuous SBOM update. However, the feedback of the projects' owners suggests the usefulness of the suggested SBOM augmentation. Future work goes in several directions, including:
\begin{compactitem}
\item Develop and evaluate tools or methodologies for automating the continuous update of augmented SBOMs to address owners' concerns about manual maintenance. 
\item Investigate how to efficiently and continuously monitor dependency changes and associated CVEs. 
\item Expand the study to include a diverse set of open- and closed-source projects. 
\item Get feedback from project owners and developers to refine the presentation and usability of augmented SBOMs. 
\end{compactitem}
In general, our preliminary results support the viability of this direction for future research.

\balance 
\bibliographystyle{ACM-Reference-Format}
\bibliography{main}


\begin{thebibliography}{22}


\ifx \showCODEN    \undefined \def \showCODEN     #1{\unskip}     \fi
\ifx \showDOI      \undefined \def \showDOI       #1{#1}\fi
\ifx \showISBNx    \undefined \def \showISBNx     #1{\unskip}     \fi
\ifx \showISBNxiii \undefined \def \showISBNxiii  #1{\unskip}     \fi
\ifx \showISSN     \undefined \def \showISSN      #1{\unskip}     \fi
\ifx \showLCCN     \undefined \def \showLCCN      #1{\unskip}     \fi
\ifx \shownote     \undefined \def \shownote      #1{#1}          \fi
\ifx \showarticletitle \undefined \def \showarticletitle #1{#1}   \fi
\ifx \showURL      \undefined \def \showURL       {\relax}        \fi
\providecommand\bibfield[2]{#2}
\providecommand\bibinfo[2]{#2}
\providecommand\natexlab[1]{#1}
\providecommand\showeprint[2][]{arXiv:#2}

\bibitem[{Anchore}(2020)]%
        {grype}
\bibfield{author}{\bibinfo{person}{{Anchore}}.}
  \bibinfo{year}{2020}\natexlab{}.
\newblock \bibinfo{booktitle}{\emph{{Grype: A vulnerability scanner for
  container images and filesystems}}}.
\newblock
\newblock
\shownote{Available at \url{https://github.com/anchore/grype} (Last Accessed:
  January 16, 2025)}.


\bibitem[Chaora et~al\mbox{.}(2023)]%
        {Chaora:2023}
\bibfield{author}{\bibinfo{person}{Anesu Chaora}, \bibinfo{person}{Nathan
  Ensmenger}, {and} \bibinfo{person}{L~Jean Camp}.}
  \bibinfo{year}{2023}\natexlab{}.
\newblock \showarticletitle{Discourse, Challenges, and Prospects Around the
  Adoption and Dissemination of Software Bills of Materials (SBOMs)}. In
  \bibinfo{booktitle}{\emph{2023 IEEE International Symposium on Technology and
  Society (ISTAS)}}. \bibinfo{publisher}{IEEE}, \bibinfo{pages}{1--4}.
\newblock
\urldef\tempurl%
\url{https://doi.org/10.1109/ISTAS57930.2023.10305922}
\showDOI{\tempurl}


\bibitem[Chen et~al\mbox{.}(2020)]%
        {ChenSS020}
\bibfield{author}{\bibinfo{person}{Yang Chen}, \bibinfo{person}{Andrew~E.
  Santosa}, \bibinfo{person}{Asankhaya Sharma}, {and} \bibinfo{person}{David
  Lo}.} \bibinfo{year}{2020}\natexlab{}.
\newblock \showarticletitle{Automated identification of libraries from
  vulnerability data}. In \bibinfo{booktitle}{\emph{2020 IEEE/ACM 42nd
  International Conference on Software Engineering: Software Engineering in
  Practice (ICSE-SEIP)}}. \bibinfo{publisher}{{ACM}}, \bibinfo{pages}{90--99}.
\newblock


\bibitem[Chisnall(1993)]%
        {Oppenheim:1992}
\bibfield{author}{\bibinfo{person}{Peter~M Chisnall}.}
  \bibinfo{year}{1993}\natexlab{}.
\newblock \showarticletitle{Questionnaire design, interviewing and attitude
  measurement}.
\newblock \bibinfo{journal}{\emph{Journal of the Market Research Society}}
  \bibinfo{volume}{35}, \bibinfo{number}{4} (\bibinfo{year}{1993}),
  \bibinfo{pages}{392--393}.
\newblock


\bibitem[Cybersecurity and (CISA)(2023)]%
        {vex}
\bibfield{author}{\bibinfo{person}{Cybersecurity} {and}
  \bibinfo{person}{Infrastructure Security~Agency (CISA)}.}
  \bibinfo{year}{2023}\natexlab{}.
\newblock \bibinfo{title}{{Minimum Requirements for Vulnerability
  Exploitability eXchange (VEX)}}.
\newblock
\newblock
\newblock
\shownote{Available at
  \url{https://www.cisa.gov/sites/default/files/2023-04/minimum-requirements-for-vex-508c.pdf}
  (Last Accessed: January 16, 2025)}.


\bibitem[Cybersecurity and (CISA)(2024)]%
        {CISA:framing}
\bibfield{author}{\bibinfo{person}{Cybersecurity} {and}
  \bibinfo{person}{Infrastructure Security~Agency (CISA)}.}
  \bibinfo{year}{2024}\natexlab{}.
\newblock \bibinfo{booktitle}{\emph{Framing Software Component Transparency:
  Establishing a Common Software Bill of Materials (SBOM)}
  (\bibinfo{edition}{3rd} ed.)}.
\newblock
\newblock
\shownote{Available at
  \url{https://www.cisa.gov/resources-tools/resources/framing-software-component-transparency-2024}
  (Last Accessed: January 16, 2025)}.


\bibitem[{GitHub Inc.}(2019)]%
        {dependabot}
\bibfield{author}{\bibinfo{person}{{GitHub Inc.}}}
  \bibinfo{year}{2019}\natexlab{}.
\newblock \bibinfo{booktitle}{\emph{{Dependabot: Automated dependency updates
  built into GitHub}}}.
\newblock
\newblock
\shownote{Available at \url{https://github.com/dependabot} (Last Accessed:
  January 16, 2025)}.


\bibitem[{Google Inc.}(2022)]%
        {osvscanner}
\bibfield{author}{\bibinfo{person}{{Google Inc.}}}
  \bibinfo{year}{2022}\natexlab{}.
\newblock \bibinfo{booktitle}{\emph{{OSV Scanner: Vulnerability scanner written
  in Go which uses the data provided by OSV}}}.
\newblock
\newblock
\shownote{Available at \url{https://github.com/google/osv-scanner} (Last
  Accessed: January 16, 2025)}.


\bibitem[Governmen(2021)]%
        {USA:ExecutiveOrder}
\bibfield{author}{\bibinfo{person}{The United States~Federal Governmen}.}
  \bibinfo{year}{2021}\natexlab{}.
\newblock \bibinfo{booktitle}{\emph{{Executive Order on Improving the Nation's
  Cybersecurity | The White House}}}.
\newblock
\newblock
\shownote{Available at
  \url{https://www.whitehouse.gov/briefing-room/presidential-actions/2021/05/12/executive-order-on-improving-the-nations-cybersecurity/}
  (Last Accessed: January 16, 2025)}.


\bibitem[Kloeg et~al\mbox{.}(2024)]%
        {Kloeg:2024}
\bibfield{author}{\bibinfo{person}{Berend Kloeg}, \bibinfo{person}{Aaron~Yi
  Ding}, \bibinfo{person}{Sjoerd Pellegrom}, {and} \bibinfo{person}{Yury
  Zhauniarovich}.} \bibinfo{year}{2024}\natexlab{}.
\newblock \showarticletitle{Charting the Path to {SBOM} Adoption: A Business
  Stakeholder-Centric Approach}. In \bibinfo{booktitle}{\emph{ASIA CCS '24:
  Proceedings of the 19th ACM Asia Conference on Computer and Communications
  Security}}. \bibinfo{publisher}{ACM}, \bibinfo{pages}{1770–1783}.
\newblock
\urldef\tempurl%
\url{https://doi.org/10.1145/3634737.3637659}
\showDOI{\tempurl}


\bibitem[Nocera et~al\mbox{.}(2024)]%
        {Nocera:2024}
\bibfield{author}{\bibinfo{person}{Sabato Nocera},
  \bibinfo{person}{Massimiliano Di~Penta}, \bibinfo{person}{Rita Francese},
  \bibinfo{person}{Simone Romano}, {and} \bibinfo{person}{Giuseppe
  Scanniello}.} \bibinfo{year}{2024}\natexlab{}.
\newblock \showarticletitle{{If it's not SBOM, then what? How Italian
  Practitioners Manage the Software Supply Chain}}. In
  \bibinfo{booktitle}{\emph{2024 IEEE International Conference on Software
  Maintenance and Evolution (ICSME)}}. \bibinfo{publisher}{IEEE},
  \bibinfo{pages}{730--740}.
\newblock
\urldef\tempurl%
\url{https://doi.org/10.1109/ICSME58944.2024.00077}
\showDOI{\tempurl}


\bibitem[Nocera et~al\mbox{.}(2023)]%
        {Nocera:2023}
\bibfield{author}{\bibinfo{person}{Sabato Nocera}, \bibinfo{person}{Simone
  Romano}, \bibinfo{person}{Massimiliano Di~Penta}, \bibinfo{person}{Rita
  Francese}, {and} \bibinfo{person}{Giuseppe Scanniello}.}
  \bibinfo{year}{2023}\natexlab{}.
\newblock \showarticletitle{Software Bill of Materials Adoption: A Mining Study
  from GitHub}. In \bibinfo{booktitle}{\emph{2023 IEEE International Conference
  on Software Maintenance and Evolution (ICSME)}}. IEEE,
  \bibinfo{pages}{39--49}.
\newblock
\urldef\tempurl%
\url{https://doi.org/10.1109/ICSME58846.2023.00016}
\showDOI{\tempurl}


\bibitem[{openvex}(2023)]%
        {vexctl}
\bibfield{author}{\bibinfo{person}{{openvex}}.}
  \bibinfo{year}{2023}\natexlab{}.
\newblock \bibinfo{booktitle}{\emph{{vexctl: A tool to make VEX work}}}.
\newblock
\newblock
\shownote{Available at \url{https://github.com/openvex/vexctl} (Last Accessed:
  January 16, 2025)}.


\bibitem[Parliament and of~the European~Union(2024)]%
        {EU:CRA}
\bibfield{author}{\bibinfo{person}{The~European Parliament} {and}
  \bibinfo{person}{The~Council of~the European~Union}.}
  \bibinfo{year}{2024}\natexlab{}.
\newblock \bibinfo{booktitle}{\emph{Cyber Resilience Act}}.
\newblock
\newblock
\shownote{Available at
  \url{https://eur-lex.europa.eu/legal-content/EN/TXT/?uri=CELEX\%3A32024R2847}
  (Last Accessed: January 16, 2025)}.


\bibitem[{Sourcegraph Inc.}(2013)]%
        {sourcegraph}
\bibfield{author}{\bibinfo{person}{{Sourcegraph Inc.}}}
  \bibinfo{year}{2013}\natexlab{}.
\newblock \bibinfo{booktitle}{\emph{{Code Intelligence for Untangling Big,
  Messy Databases}}}.
\newblock
\newblock
\shownote{Available at \url{https://sourcegraph.com} (Last Accessed: January
  16, 2025)}.


\bibitem[Stalnaker et~al\mbox{.}(2024)]%
        {Stalnaker:2023}
\bibfield{author}{\bibinfo{person}{Trevor Stalnaker}, \bibinfo{person}{Nathan
  Wintersgill}, \bibinfo{person}{Oscar Chaparro}, \bibinfo{person}{Massimiliano
  Di~Penta}, \bibinfo{person}{Daniel~M German}, {and} \bibinfo{person}{Denys
  Poshyvanyk}.} \bibinfo{year}{2024}\natexlab{}.
\newblock \showarticletitle{{BOMs} Away! Inside the Minds of Stakeholders: A
  Comprehensive Study of Bills of Materials for Software Systems}. In
  \bibinfo{booktitle}{\emph{2024 IEEE/ACM 46th International Conference on
  Software Engineering (ICSE)}}. \bibinfo{publisher}{ACM},
  \bibinfo{pages}{44:1--44:13}.
\newblock
\urldef\tempurl%
\url{https://doi.org/10.1145/3597503.3623347}
\showDOI{\tempurl}


\bibitem[Tal(2023)]%
        {sca1}
\bibfield{author}{\bibinfo{person}{Liran Tal}.}
  \bibinfo{year}{2023}\natexlab{}.
\newblock \bibinfo{booktitle}{\emph{Guide to Software Composition Analysis}}.
\newblock
\newblock
\shownote{Available at
  \url{https://snyk.io/series/open-source-security/software-composition-analysis-sca/}
  (Last Accessed: January 16, 2025)}.


\bibitem[{The Linux Foundation}(2011)]%
        {SpdxDraft:2010}
\bibfield{author}{\bibinfo{person}{{The Linux Foundation}}.}
  \bibinfo{year}{2011}\natexlab{}.
\newblock \bibinfo{booktitle}{\emph{Software Package Data eXchange (SPDX)}}.
\newblock
\newblock
\shownote{Available at \url{https://spdx.dev/use/specifications/} (Last
  Accessed: January 16, 2025)}.


\bibitem[{The Linux Foundation}(2022)]%
        {Linux:2022}
\bibfield{author}{\bibinfo{person}{{The Linux Foundation}}.}
  \bibinfo{year}{2022}\natexlab{}.
\newblock \bibinfo{booktitle}{\emph{The State of Software Bill of Materials
  ({SBOM}) and Cybersecurity Readiness}}.
\newblock
\newblock
\shownote{Available at
  \url{https://www.linuxfoundation.org/research/the-state-of-software-bill-of-materials-sbom-and-cybersecurity-readiness}
  (Last Accessed: January 16, 2025)}.


\bibitem[{The OWASP Foundation}(2018)]%
        {cyclonedx}
\bibfield{author}{\bibinfo{person}{{The OWASP Foundation}}.}
  \bibinfo{year}{2018}\natexlab{}.
\newblock \bibinfo{booktitle}{\emph{{CycloneDX}}}.
\newblock
\urldef\tempurl%
\url{https://cyclonedx.org}
\showURL{%
\tempurl}
\newblock
\shownote{Available at \url{https://cyclonedx.org}(Last Accessed: January 16,
  2025)}.


\bibitem[Wohlin et~al\mbox{.}(2012)]%
        {Wohlin:2012}
\bibfield{author}{\bibinfo{person}{Claes Wohlin}, \bibinfo{person}{Per
  Runeson}, \bibinfo{person}{Martin H{\"o}st}, \bibinfo{person}{Magnus~C
  Ohlsson}, \bibinfo{person}{Bj{\"o}rn Regnell}, {and} \bibinfo{person}{Anders
  Wessl{\'e}n}.} \bibinfo{year}{2012}\natexlab{}.
\newblock \bibinfo{booktitle}{\emph{Experimentation in Software Engineering}}.
\newblock \bibinfo{publisher}{Springer}.
\newblock


\bibitem[Xia et~al\mbox{.}(2023)]%
        {Xia:2023}
\bibfield{author}{\bibinfo{person}{Boming Xia}, \bibinfo{person}{Tingting Bi},
  \bibinfo{person}{Zhenchang Xing}, \bibinfo{person}{Qinghua Lu}, {and}
  \bibinfo{person}{Liming Zhu}.} \bibinfo{year}{2023}\natexlab{}.
\newblock \showarticletitle{An Empirical Study on Software Bill of Materials:
  Where We Stand and the Road Ahead}. In \bibinfo{booktitle}{\emph{2023
  IEEE/ACM 45th International Conference on Software Engineering (ICSE)}}.
  \bibinfo{publisher}{{IEEE}}, \bibinfo{pages}{2630--2642}.
\newblock
\urldef\tempurl%
\url{https://doi.org/10.1109/ICSE48619.2023.00219}
\showDOI{\tempurl}


\end{thebibliography}

\end{document}